\documentclass[11pt]{article}
\usepackage{amsmath,amssymb,color}

\usepackage{amsfonts}

\def\n{\hat{n}}
\newcommand{\hoch}[1]{$\, ^{#1}$}

\makeatletter
\@addtoreset{equation}{section}
\makeatother

\newcommand{\be}{\begin{equation}}
\newcommand{\ee}{\end{equation}}
\newcommand{\bea}{\setlength\arraycolsep{2pt} \begin{eqnarray}}
\newcommand{\eea}{\end{eqnarray}}
\newcommand{\nn}{\nonumber}

\def\ft#1#2{{\textstyle{\frac{\scriptstyle #1}{\scriptstyle #2} } }}
\def\fft#1#2{{\frac{#1}{#2}}}

\def\0{{\sst{(0)}}}
\def\1{{\sst{(1)}}}
\def\2{{\sst{(2)}}}
\def\3{{\sst{(3)}}}
\def\4{{\sst{(4)}}}
\def\5{{\sst{(5)}}}
\def\6{{\sst{(6)}}}
\def\7{{\sst{(7)}}}
\def\8{{\sst{(8)}}}
\def\sst#1{{\scriptscriptstyle #1}}

\def\del{{\partial}}

\def\binom#1#2{{C_{#1}^{#2}}}

\begin{document}

\begin{flushright}
\hfill{ \
MIFPA-13-03\ \ \ \ }
\end{flushright}

\vspace{25pt}
\begin{center}
{\Large {\bf Black Holes in Six-dimensional Conformal Gravity}
}

\vspace{30pt}

{\Large
H. L\"u\hoch{1}, Yi Pang\hoch{2} and C.N. Pope\hoch{2,3}
}

\vspace{10pt}

\hoch{1}{\it Department of Physics, Beijing Normal University,
Beijing 100875, China}

\vspace{10pt}

\hoch{2} {\it George P. \& Cynthia Woods Mitchell  Institute
for Fundamental Physics and Astronomy,\\
Texas A\&M University, College Station, TX 77843, USA}

\vspace{10pt}

\hoch{3}{\it DAMTP, Centre for Mathematical Sciences,
 Cambridge University,\\  Wilberforce Road, Cambridge CB3 OWA, UK}

\vspace{20pt}

\underline{ABSTRACT}
\end{center}
\vspace{15pt}

We study conformally-invariant theories of gravity in six dimensions.
In four dimensions, there is a unique such theory that is polynomial
in the curvature and its derivatives, namely Weyl-squared, and furthermore
all solutions of Einstein gravity are also solutions of the conformal theory.
By contrast, in six dimensions there are three independent
conformally-invariant polynomial terms one could consider.  There is a
unique linear combination (up to overall scale) for which Einstein metrics
are also solutions, and this specific theory forms the focus of our
attention in this paper.  We reduce the equations of motion for the most
general spherically-symmetric black hole to a single 5th-order differential
equation.  We obtain the general solution in the form of an infinite series,
characterised by 5 independent parameters, and we show how a finite
3-parameter truncation reduces to the already known Schwarzschild-AdS metric
and its conformal scaling.  We derive general results for the thermodynamics
and the first law for the full 5-parameter solutions. We also investigate
solutions in extended theories coupled to conformally-invariant matter, and
in addition we derive some general results for conserved charges in
cubic-curvature theories in arbitrary dimensions.

\thispagestyle{empty}

\pagebreak
\voffset=-40pt
\setcounter{page}{1}

\tableofcontents

\addtocontents{toc}{\protect\setcounter{tocdepth}{2}}


\newpage

\section{Introduction}

   Higher-derivative gravity theories are of interest for a variety of
reasons.  They arise naturally in string theory and M-theory,
in the form of higher-order corrections to the leading Einstein-Hilbert
term in the low-energy effective action.  In this context, the corrections
take the form of an infinite series of terms that involve derivatives of
arbitrarily high order.  There are also situations where it is of
interest to consider theories where there are just a finite number of
higher-derivative terms.  Examples include topologically massive
gravity in three dimensions \cite{desjactem,lisost}, where a
gravitational Chern-Simons term
gives a three-derivative contribution, proportional to the Cotton tensor;
New Massive Gravity in three dimensions \cite{berhohtow},
where there is a four-derivative
contribution arising from a curvature-squared term in the action; and
numerous higher-dimensional examples involving curvature-squared or
higher modifications to Einstein gravity.  Recent examples that have been
considered in four dimensions include Einstein gravity with a cosmological
constant, with an additional Weyl-squared term whose coefficient may be
tuned to give ``critical gravity'' for which the additional normally massive
spin-2 excitations around an AdS background become massless \cite{lpcritical};
 and
pure Weyl-squared conformal gravity, which has been argued to be equivalent to
Einstein gravity with a cosmological constant \cite{mald}.  In dimensions
$D\le 6$, supersymmetric extensions of certain higher-derivative theories
are also known.  These can arise because the supersymmetry is realised
off-shell, with the added higher-derivative bosonic terms being
extended to complete and independent super-invariants.  Thus, unlike
the situation in the string or M-theory effective actions, where
supersymmetry is on-shell and works order by order, requiring an infinity of
higher-order terms, in the off-shell supergravities only a finite
number of terms are required.

   In four dimensions there is a unique conformally-invariant pure
gravity theory that is polynomial in the curvature, for which the action
is given by the square of the
Weyl tensor.  It has the important feature that any Einstein
metric is also a solution of the conformal theory.  Furthermore, since
any conformal scaling of a solution is also a solution, this means that
any conformally-Einstein metric is automatically a solution of
four-dimensional conformal gravity.  This is a useful property when one
is looking for solutions to the theory, since previously-known ones
from Einstein gravity will be solutions too.  Of course, since the
equations of motion of the conformal gravity are of higher order than
those in Einstein gravity, there will exist further solutions over and
above those of Einstein gravity.  In a recent paper \cite{lopapova},
various classes of solutions in four-dimensional conformal gravity were
investigated in detail, including spherically symmetric asymptotically AdS
black holes, and black holes obeying asymptotically Lifshitz boundary
conditions.  The general spherically-symmetric asymptotically
AdS black hole solution was already known \cite{Riegert:1984zz}.
It has one additional parameter,
over and above the mass and the cosmological constant of the
Schwarzschild-AdS black hole.  This parameter can be understood as coming
from the freedom to make a (spherically-symmetric) conformal rescaling
of Schwarzschild-AdS.  It does, nevertheless, provide an interesting
extension of the usual Schwarzschild-AdS black holes, in which the
additional parameter can be interpreted as a characterisation of
massive spin-2 ``hair.'' In \cite{lopapova}, the thermodynamics
and the first law for these extended solutions was studied.
In four dimensions, charged rotating black holes \cite{Liu:2012xn}
and the generalized Plebanski solutions \cite{Mannheim:1990ya} were also
obtained. The neutral solutions are all conformal to
Einstein metrics \cite{Liu:2012xn}.

   In the present paper, we carry out some analogous investigations in
six-dimensional conformal gravity.  The situation is more complicated
in six dimensions because there is no longer a unique choice of
conformal theory.  In fact, there
is a three-parameter family of conformal gravities in six dimensions
that have actions polynomial in the curvature and its derivatives
(see \cite{
mets,lupapo}), described by the action
$I=\beta_1\, I_1+\beta_2 I_2 + \beta_3 I_3$, where\footnote{We use the same
conventions as in \cite{bcn}.}
\bea
I_1 &=& C_{\mu\rho\sigma\nu} C^{\mu\alpha\beta\nu}
 C_{\alpha}{}^{\rho\sigma}{}_\beta\,,\nn\\
I_2 &=& C_{\mu\nu\rho\sigma} C^{\rho\sigma\alpha\beta}
           C_{\alpha\beta}{}^{\mu\nu}\,,\nn\\
I_3 &=& C_{\mu\rho\sigma\lambda}\Big(\delta^\mu_\nu\, \Box +
    4R^\mu{}_\nu - \fft65 R\, \delta^\mu_\nu\Big) C^{\nu\rho\sigma\lambda}
   + \nabla_\mu J^\mu\,,\nn\\
J^{\mu}&=&4R_{\mu}^{~\lambda\rho\sigma}\nabla^{\nu}R_{\nu\lambda\rho\sigma}+
3R^{\nu\lambda\rho\sigma}\nabla_{\mu}R_{\nu\lambda\rho\sigma}
-5R^{\nu\lambda}\nabla_{\mu}R_{\nu\lambda}
+\ft12R\nabla_{\mu}R\cr
&&-R_{\mu}^{~\nu}\nabla_{\nu}R+2R^{\nu\lambda}\nabla_{\nu}R_{\lambda\mu},
\eea
and the coefficients $\beta_1$, $\beta_2$ and $\beta_3$ are arbitrary.
In general, Einstein metrics will not be solutions of the theory, except in
the special case where $\beta_1= 4\beta_2=-12\beta_3$.  In particular,
with this choice of parameters the theory
allows Schwarzschild-AdS black
holes as solutions, and this has the advantage that at least some
explicit spherically-symmetric solutions are available for investigation.

   Accordingly, we shall consider the Lagrangian
\begin{eqnarray}
e^{-1} {\cal L}_{\rm conf} &=&\beta ( 4 I_1 + I_2 - \ft13 I_3)\cr
&=&\beta
\Big(RR^{\mu\nu}R_{\mu\nu}-\ft{3}{25} R^3-2R^{\mu\nu}R^{\rho\sigma}
R_{\mu\rho\nu\sigma} -R^{\mu\nu}\Box R_{\mu\nu}+\ft{3}{10}R\Box
R\Big)\nn\\
&& +\hbox{total derivative}\,,\label{d6conflag}
\end{eqnarray}
The equations of motion of this system are given by
\be
E_{\mu\nu}\equiv E^{(1)}_{\mu\nu}-\ft3{25}E^{(2)}_{\mu\nu}-2E^{(3)}_{\mu\nu}
-E^{(4)}_{\mu\nu}+\ft3{10}E^{(5)}_{\mu\nu}=0\,,\label{d6eom}
\ee
where the individual contributions $E^{(n)}_{\mu\nu}$ coming from the
variation of each term in (\ref{d6conflag}) are given in Appendix B.

   In section 2, we study the equations of motion for spherically-symmetric
black hole solutions.  These can be reduced to a 5th-order ordinary differential
equation for a single undetermined metric function.  As mentioned above,
the Schwarzschild-AdS metric of six-dimensional Einstein gravity is
a solution, and furthermore, any conformal scaling is also a solution.
This provides us with an explicit three-parameter family of
spherically-symmetric black hole solutions, but, unlike the situation
in four-dimensional conformal gravity, this does not exhaust the
space of solutions, which should be characterised by a total of five
parameters.  We have not been able to construct the most general such
solutions explicitly, but we have constructed it as
an infinite series expansion for the metric function, with explicit
expansion coefficients.

  In section 3, we use the Noether procedure to construct a conserved
charge which, when integrated over a compact spatial surface at infinity,
provides an expression for the mass of the black hole.  Only the first
few terms in our series expansion for the metric function contribute in
this asymptotic formula, and so we are able to obtain an explicit
expression for the mass of the general five-parameter solution.  The same
conserved charge, when integrated over the horizon, yields the expression
for the product $TS$ of the temperature and the entropy.  Furthermore,
the temperature itself can be calculated via a computation of the
surface gravity.  By this means, we are able to obtain explicit
expressions for the temperature and entropy of the exact three-parameter
family of black holes whose expression can be given in closed form.

  In section 4, using the general methods developed by Wald
\cite{Wald:1993nt,Iyer:1994ys}, we use the conserved charge mentioned above to derive the
first law of thermodynamics for the general five-parameter spherically-symmetric
black holes.  We also derive a Smarr-type formula for these solutions.

  In section 5, we discuss extensions of the conformal gravity theory in
which conformally-invariant ``matter'' is added also.  In particular, this
can include a 2-form potential, and also an electromagnetic field
whose field strength couples quadratically to the Weyl tensor.  In section 6 we
discuss various further explicit solutions of conformal gravity and these
conformal matter extensions.

   In section 7, we give a general discussion of the calculation of
conserved charges in curvature-cubed theories of gravity in arbitrary
dimensions, using the general conformal methods developed by Ashtekar,
Magnon and Das (AMD) \cite{Ashtekar:1984zz,ashdas}.  In section 8 we
discuss {\it tricritical gravity} in six dimensions, which was
first constructed in \cite{lupapo}.\footnote{The unitarity problem and
consistent truncation of ghost-like logarithmic modes in multi-critical
gravity theories were
studied in \cite{Bhmrz,Nut,knv}.  It was shown that at the level of
the free theory, in special cases they could admit a unitary subspace.
However, as pointed out later by \cite{ap}, the analysis carried out at
for the free theory is invalid at the non-linear level, and the
would-be unitary subspace suffers from a linearisation instability and
is absent in the full non-linear theory. Including the ghost-like
logarithmic modes seems to be indispensable for the consistency of the
theory. As a consequence, these multi-critical gravity theories were
conjectured to be the gravity duals of multi-rank logarithmic CFTs.}
This is obtained
by appending an Einstein-Hilbert term, a cosmological term, and a
Weyl-squared term to the conformal theory that we have been studying in this
paper.  The coefficients of the additional terms are tuned so that the
additional spin-2 modes around an AdS background, which are
generically massive,  become massless.  We include a discussion of
consistent boundary conditions that can be imposed in this theory.  Following
the conclusions in section 9, we then include two appendices. In appendix A,
we review the derivation of a useful necessary condition
\cite{koneto,gonu} that must
be satisfied by any metric that is conformal to an Einstein metric.  This
provides a valuable tool when investigating whether a given solution in
the conformal gravity might be ``new,'' as opposed to merely being a
conformal scaling of a previously-known Einstein metric.  Finally,
in appendix B, we give expressions for the contributions to the
field equations that result from the various six-derivative terms that
arise in the six-dimensional theories that we are considering.

\section{Static Black Hole Solutions}

We shall consider the ansatz for static solutions of the form
\begin{equation}
ds^2=-f\, dt^2 + \fft{dr^2}{f} + r^2 d\Omega_{4,k}^2\,,\label{sphsym}
\end{equation}
where for $k=1$, $-1$ or 0 the metric $d\Omega_{4,k}^2$ describes a
unit 4-sphere, hyperbolic 4-space or the 4-torus respectively, and $f$ is
a function of $r$.  (The metric functions in $g_{tt}$ and $g_{rr}$ can be taken
to be inversely related, as we have done here, by using the conformal
symmetry.)  Since we shall typically be concentrating on the $k=1$ case
we shall commonly refer to the metric as being ``spherically symmetric,''
even when $k$ is unspecified.
Substituting the ansatz into the
equations of motion (\ref{d6eom}), we find that all the equations are satisfied
provided that the equation $E_{rr}=0$ is satisfied.  This gives rise to a
5th-order differential
equation for the function $f(r)$. (Analogous solutions for an action
using just $I_1$ and $I_2$ were obtained in \cite{oliray}.)

It is in fact possible to exploit the conformal symmetry of the problem to
obtain a simpler parameterisation.  Passing to the conformally-related
metric $d\hat s^2$, and introducing a new radial coordinate $\rho$ and metric
function $h(\rho)$ defined by
\be
d\hat{s}^2=r^{-2}ds^2,\qquad \rho=1/r,\qquad h(\rho)=r^{-2}f(r)\,,
\ee
we obtain
\be
d\hat{s}^2=-h(\rho)\, dt^2 + \fft{d\rho^2}{h(\rho)} + d\Omega_{4,k}^2\,.
\ee
Now in the new metric, the equations of motion imply simply
\bea
&&-216 k^3+42 k h''^2+6 h''^3-84 k h'h^{(3)}-18 h'h''h^{(3)}+5 h(h^{(3)})^2\nn\\
&&+20 h'^2h^{(4)}-10h h''h^{(4)}+10hh' h^{(5)} =0,\label{sphsym2}
\eea
where a prime means a derivative with respect to $\rho$, and $h^{(n)}$ denotes
the $n$'th derivative of $h$.

  If Eqn. (\ref{sphsym2}) is differentiated once more,
it yields a rather simple 6th order equation\footnote{In the case of
four-dimensional conformal gravity,
the analogous equation that results from differentiating the 3rd-order
equation for $h$ is simply $h^{(4)}=0$, showing that the
general spherically-symmetric static solution of four-dimensional
conformal gravity is given by a third-order polynomial.}
\be
10hh^{(6)}+30h'h^{(5)}+12h''h^{(4)}-13(h^{(3)})^2-84kh^{(4)}=0
\label{sphsym3}\,.
\ee

  Using on Eqs.(\ref{sphsym2}) and (\ref{sphsym3}), we can obtain the general
 spherically-symmetric solution as a series expansion of the form
\be
h(\rho)=\sum_{n\geq0}\frac{b_n}{n!}\, \rho^n\,,\label{gs}
\ee
where $\{b_0,b_1,b_2,b_3,b_4\}$ are free parameters, while $b_5$ and
$b_n$, $(n\geq6)$ are determined by
\bea
&&6 b_2^3-18 b_1 b_2 b_3+5 b_0 b_3^2+42k b_2^2 -84k b_1 b_4 -216k^3+20b_1^2 b_4-10 b_0 b_2 b_4+10 b_0 b_1 b_5=0,\cr
&&10b_0b_{2n}+\sum_{m=1}^{n-1}\alpha (2n,m)b_{m}b_{2n-m}+\beta(2n,n)b_{n}^2-84kb_{2n-2}=0,\quad n\geq3, \cr
&&10b_0b_{2n+1}+\sum_{m=1}^{n}\alpha (2n+1,m)b_{m}b_{2n+1-m}-84kb_{2n-1}=0\,,
\quad n\geq3,
\eea
with the coefficients $\alpha(n,m)$ and $\beta(2n,n)$ given by
\bea
&&\alpha(n,m)=2\binom{n-6}{m-6}+30\binom{n-6}{m-5}+
  12\binom{n-6}{m-4}-26\binom{n-6}{m-3}+12\binom{n-6}{m-2}
+30\binom{n-6}{m-1}+10\binom{n-6}{m}\,,\cr
&&\beta(2n,n)=10\binom{2n-6}{n}+30\binom{2n-6}{n-1}+
12\binom{2n-6}{n-2}-13\binom{2n-6}{n-3}\,.
\eea
Here $\binom{n}{k}=n!/(k! \, (n-k)!)$ is the binomial coefficient, and
it is understood that the factorial of a
negative integer is infinity. Among the five-parameter
general solution Eq.(\ref{gs}) there are black hole solutions.

   In terms of the previous spherically-symmetric ansatz Eq.(\ref{sphsym}),
the five-parameter solution takes the form
\be
a(r)=f(r)=r^2\Big(a_0+ \fft{a_1}{r} + \fft{a_2}{r^2} + \fft{a_3}{r^3} +
\fft{a_4}{r^4} + \fft{a_5}{r^5}+\sum_{n\geq6}\frac{a_n}{r^n}\Big).
\ee
Here we use parameters $a_n=b_n/n!$ for later convenience.

We find that there exists a three-parameter subset of solutions that
corresponds to a finite truncation of the five-parameter general solutions.
In terms of the usual parametrization Eq.(\ref{sphsym}), it is given by
\begin{equation}
f=r^2 \Big(a_0 + \fft{a_1}{r} + \fft{a_2}{r^2} + \fft{a_3}{r^3} +
\fft{a_4}{r^4} + \fft{a_5}{r^5}\Big)\,,\label{ffn}
\end{equation}
where
\begin{equation}
a_1=\fft{a_4(a_4^3+50ka_5^2)}{125a_5^3}\,,\quad
a_2=k+\fft{2a_4^3}{25a_5^2}\,,\quad a_3=\fft{2a_4^2}{5a_5}\,,\quad
a_n=0\ \hbox{for}\ n\ge 6\,.\label{arels}
\end{equation}

   In fact this three-parameter subset of the general solutions admits of a
very simple interpretation.  As we already noted, any solution of the Einstein
equations is also a solution of the specific conformally-invariant theory
we are considering here.  Furthermore, any conformal scaling of an Einstein
metric will also be a solution.  The solutions given by (\ref{ffn}) and
(\ref{arels}) are in fact precisely the family of conformal scalings of
the Schwarzschild-AdS metric that can be cast within the form of the ansatz
(\ref{sphsym}).  To see this, we start from the
Schwarzschild-AdS metric in the standard form
\be ds_{\rm{SAdS}}^2= -\Big(k+ y^2/L^2 -\fft{m}{y^3}\Big)\,
dt^2 + \Big(k+y^2/L^2 -\fft{m}{y^3}\Big)^{-1}\,
dy^2 + y^2\,d\Omega_{4,k}^2\,,\label{schadsbh}
\ee
which satisfies $R_{\mu\nu}=-5 L^2\, g_{\mu\nu}$.
The metrics (\ref{sphsym}) with $f$ given by (\ref{ffn}) and (\ref{arels})
are conformally related, with $ds_{\rm SAdS}^2 = \Omega^2\, ds^2$, where
\bea
&&\Omega^2= \fft{1}{(cr+1)^2}\,,\qquad y=\frac{r}{1+cr},\cr
&&a_0=c^2k+\frac{1}{L^2}-c^5m,\quad a_1=2ck-5c^4m,\quad a_2=k-10c^3m,\cr
&&a_3=-10c^2m,\quad a_4=-5c m,\quad a_5=-m.\label{cftrans}
\eea
The ``thermalized vacuum'' corresponds to solutions with $\mu=0$
(see \cite{lopapova} for the analogous discussion in four-dimensional
conformal gravity). The thermodynamic quantities for the
 Schwarzschild-AdS black hole in six-dimensional conformal gravity
are given by
\be
E=-96\beta\frac{m}{L^4},\quad T=\frac{5m-2k y^3_{+}}{4\pi y^4_{+}},
\quad S=-96\pi\beta(\frac{y^4_{+}}{L^4}-k)\,.\label{ETS}
\ee
These quantities satisfy the first law of thermodynamics
\begin{equation}
dE = T dS\,.\label{fl0}
\end{equation}

\subsection{Spherically-symmetric solutions that are not conformally Einstein}

In \cite{lopapova}, it was shown that
the general spherically-symmetric solution of four-dimensional conformal
gravity is
conformal to the Schwarzschild-AdS (dS) metric.  By contrast, we find that
the general five-parameter solutions given in Eq.(\ref{gs}) are not
conformal to any Einstein metric.  To see this, let us suppose
that $e^{2\phi}d\hat{s}^2$ was in fact an Einstein metric.
By using the necessary condition for a metric to be conformally
Einstein in six dimensions \cite{gonu} (see Appendix A), we find
$\phi$ must be a function of $\rho$ and that
\be
\phi'=\frac{h'''}{3(2k-h'')}\,.
\ee
Combining this equality with the requirement that $e^{2\phi}d\hat{s}^2$ be
an Einstein metric implies that $h$ should satisfy
\be
3h''h^{(4)}-2(h^{(3)})^2-6kh^{(4)}=0\,,
\ee
which then implies that $h$ is a certain 5th-order polynomial in $\rho$.
Substituting back into the equations of motion for conformal gravity
then leads us back to the closed-form three-parameter solution given by
(\ref{ffn}) and (\ref{arels}).  Thus, we have proved that
the general spherically-symmetric solution of
six-dimensional conformal gravity is not conformally Einstein.

\section{Energy of AdS Black Holes in D=6 Conformal Gravity}

To calculate the energy of the black hole solutions in Eq.(\ref{gs}),
we start from the conformally invariant Lagrangian in Eq.(\ref{d6conflag}),
 and
derive the Noether charge associated with the Killing vector $\xi^{\mu}$.
We consider the variation of Lagrangian 6-form induced by $\xi^{\mu}$,
\be
  {\cal L}_{\xi}L=E^{\alpha\beta}{\cal L}_{\xi}g_{\alpha\beta}+
d\Theta(g_{\alpha\beta},{\cal L}_{\xi}g_{\alpha\beta})\,,
\ee
where $E^{\alpha\beta}$ represents the equations of motion.
When $E^{\alpha\beta}=0$ is satisfied, then
using the identity
\begin{equation}
    {\cal L}_{\xi}=di_{\xi}+i_{\xi}d,\label{id1}
\end{equation}
for the Lie derivative of a differential form,
we find a conserved current defined by
\begin{equation}
    J=\Theta-i_{\xi}L,\quad  dJ=0\Rightarrow J=dQ[\xi].\label{Qcharge}
\end{equation}
Explicitly, in six-dimensional conformal gravity,
the conserved charge is a 4-form
\be
Q[\xi]=\frac{1}{2!4!}\int\epsilon_{\alpha\beta\mu\nu\lambda\rho}
Q^{\alpha\beta}dx^{\mu}\wedge dx^{\nu}\wedge dx^{\lambda}\wedge dx^{\rho}\,,
\ee
which consists of two parts, $Q^{\alpha\beta}_1$+$Q^{\alpha\beta}_2$.
$Q^{\alpha\beta}_1$ and $Q^{\alpha\beta}_2$ depend on
$\nabla_{\mu}\xi_{\nu}$ and $\xi_{\mu}$ respectively:
\bea
&&Q_1^{\alpha\beta}=X^{\alpha\beta\mu\nu}\nabla_{\mu}\xi_{\nu},\cr
&&X^{\alpha\beta\mu\nu}=-\beta\biggl(24C^{[\alpha~|\nu|}_{~~\lambda~~\rho}C^{\beta]\lambda\mu\rho}
-6C^{[\alpha}_{~~\lambda\rho\sigma}g^{\beta][\nu}C^{\mu]\lambda\rho\sigma}
+\frac35C^{\lambda\rho}_{~~\sigma\delta}C^{\sigma\delta}_{~~\lambda\rho}g^{\alpha[\mu}g^{\nu]\beta}\biggr)\cr
&&\qquad\qquad-\beta\biggr(6C^{\alpha\beta}_{~~\lambda\rho}C^{\lambda\rho\mu\nu}
-2C^{[\alpha}_{~\lambda\rho\sigma}g^{\beta][\nu}C^{\mu]\lambda\rho\sigma}
+\frac{1}{5}C^{\lambda\rho}_{~~\sigma\delta}C^{\sigma\delta}_{~~\lambda\rho}g^{\alpha[\mu}g^{\nu]\beta}\biggr)\cr
&&\qquad\qquad+\ft23\beta\biggl(2\Box C^{\alpha\beta\mu\nu}
+4C^{[\alpha}_{~~\lambda\rho\sigma}g^{\beta][\nu}C^{\mu]\lambda\rho\sigma}+R^{\lambda[\alpha}C_{\lambda}^{~\beta]\mu\nu}
+R^{\lambda[\mu}C_{\lambda}^{~\nu]\alpha\beta}\cr
&&\qquad\qquad-R^{\lambda\rho}C_{\lambda~\rho}^{~\alpha~[\mu}g^{\nu]\beta}
+R^{\lambda\rho}C_{\lambda~\rho}^{~\beta~[\mu}g^{\nu]\alpha}\biggr) -\ft{2}{5}\beta\biggl(4RC^{\alpha\beta\mu\nu}
+C^{\lambda\rho}_{~~\sigma\delta}C^{\sigma\delta}_{~~\lambda\rho}g^{\alpha[\mu}g^{\nu]\beta}\biggr);\cr
&&Q_2^{\alpha\beta}=\xi_{\nu}\nabla_{\mu}X^{\alpha\beta\mu\nu}+\ft23\beta\xi^{[\alpha}J^{\beta]}-\ft23\beta\biggl(
2\xi^{\lambda}C^{[\beta}_{~~\nu\rho\sigma}\nabla^{\alpha]}R_{\lambda}^{~\nu\rho\sigma}
-2\xi^{\lambda}C^{[\alpha}_{~~\nu\rho\sigma}\nabla_{\lambda}R^{\beta]\nu\rho\sigma}\cr
&&\qquad\qquad-2\xi_{\lambda}C^{\lambda\nu\rho\sigma}\nabla^{[\alpha}R^{\beta]}_{~~\nu\rho\sigma}
-2\xi^{\lambda}R_{\lambda\nu\rho\sigma}\nabla^{[\alpha}C^{\beta]\nu\rho\sigma}
+2\xi^{\lambda}\nabla_{\lambda}C^{[\alpha}_{~~\nu\rho\sigma}R^{\beta]\nu\rho\sigma}\cr
&&\qquad\qquad +2\xi^{\lambda}\nabla^{[\alpha}C_{\lambda\nu\rho\sigma}R^{\beta]\nu\rho\sigma}
-C^{\lambda\rho\sigma\delta}\xi^{[\alpha}\nabla^{\beta]}
C_{\lambda\rho\sigma\delta}\biggr)\,.\label{qcharge}
\eea
When evaluated at infinity, $Q[\xi]$ gives the mass of black hole solutions
in Eq.(\ref{gs})
\be
E=\beta V(\Omega_k)\biggl(96a_0^2a_5-\ft{12}{25} (17 k^2a_1 - 14k a_1 a_2-3a_1a_2^2 + 54k a_0 a_3 +5a_1^2a_3+46 a_0a_2 a_3  -
   60 a_0 a_1 a_4 )\biggr)\,,\label{bhmass}
\ee
after using the on-shell relations among the $a_i$s.  It should be emphasised
that this expression for the mass is valid for the full five-parameter
family of solutions.

When evaluated on the horizon, $Q[\xi]$ is equal to $TS$.
Since $\nabla_{\mu}\xi_{\nu}=\kappa\epsilon_{\mu\nu}$, where $\kappa$ is
the surface gravity on the horizon, $\epsilon_{\mu\nu}$ is the bi-normal
vector horizon normalized
to satisfy $\epsilon_{\mu\nu}\epsilon^{\mu\nu}=-2$ and the
Killing vector $\xi$ vanishes on the horizon, it follows that
the entropy formula can be simplified to give
\be
S=\pi\int_{\cal H}X^{\alpha\beta\mu\nu}\epsilon_{\alpha\beta}
\epsilon_{\mu\nu}d\Sigma\,,
\ee
where $X^{\alpha\beta\mu\nu}$ is defined in the first line of
Eq.(\ref{qcharge}). Explicit calculation for the three-parameter
closed-form solutions (\ref{ffn}) and (\ref{arels}) leads to
\be
S=-\frac{\beta96\pi V(\Omega_k)(5a_5+a_4r_{+})^3(125a_5^3+15a_4^2a_5r^2_{+}+a_4^3r^3_{+
}+75a_4a_5^2r_{+}+250a_5^2r^3_{+})}{15625a_5^4r^6_{+}}\,,\label{entropy}
\ee
where $r_{+}$ is the largest positive root of $f(r)=0$, i.e. it is
the radius of outer horizon. The temperature is given by
\be
T=-\frac{(5a_5+a_4r_{+})(125a_5^3+75a_4a_5^2r_{+}+15a_4^2a_5r_{+}^2+
a_4^3r_{+}^3+50a_5^2r_{+}^3)}{500\pi a_5^3r_{+}^4}\,.\label{temperature}
\ee
By using the parameter relations in Eq.(\ref{cftrans}), it can be shown
that the entropy and temperature of the three-parameter
black holes in six-dimensional conformal gravity are equal to those of the
conformally-related Schwarzschild-AdS black hole. In other
words, the entropy and temperature are conformal invariants, as is also
the case in four dimensions \cite{lopapova}.\footnote{It should be emphasised
that the expression for the mass of the black holes, given by (\ref{bhmass}),
is valid for the general five-parameter solutions, since it is evaluated
on a surface at infinity where only the leading orders in the radial fall-off
contribute.  By contrast, the entropy (\ref{entropy}) and temperature
(\ref{temperature}) are evaluated on the horizon, and so without having
closed-form expressions for the general solutions, these can only be
evaluated explicitly for the three-parameter closed-form truncation.}
This is related to the fact that conformal factor $\Omega^2$
in (\ref{cftrans}) is regular on the horizon and the near-horizon
geometry is preserved.  By contrast, the asymptotic region of the
Schwarzschild black hole is altered by the conformal transformation and
hence the Schwarzschild black hole energy in (\ref{ETS})
becomes (\ref{bhmass}).  It follows that the first law of
thermodynamics (\ref{fl0}) of the Schwarzschild black hole no longer
applies. The three-parameter black hole is a
globally distinct spacetime even though it is locally conformal to the
Schwarzschild black hole. We shall derive the first law of
thermodynamics in the next section.

  If we define the Helmholtz energy to be $F=-TI_{E}$,
where $I_{E}$ is the Euclidean action, then we find that
\be
F=E-TS.
\ee
A simple way to see this is to calculate the Euclidean action of the
conformally-related Schwarzschild-AdS metric with
$y\in[y_{+}, 1/c$] (see Eq.(\ref{cftrans})).

\section{Black Hole Thermodynamics}

In the previous section, we derived the conserved quantities in six-dimensional
conformal gravity by the Noether method.
The expressions for the entropy and temperature of the general spherical
solution are given by
\be
T=-\frac{h'(\rho_{+})}{4\pi},\qquad S=\frac{4\pi\beta V(\Omega_k)}{25}\Big(204-84 h''(\rho_{+})-9 h''(\rho_{+})^2+10 h'(\rho_{+}) h^{(3)}(\rho_{+})\Big).
\ee
From Eq.(\ref{Qcharge}), one can see that
\be
dQ=-i_{\xi}L.
\ee
Evaluating this equation in the region bounded by horizon and infinity
just gives
\be
F=E-TS.
\ee
To study the first law of thermodynamics, we follow the construction
of \cite{Wald:1993nt,Iyer:1994ys}.  We do this by considering the difference
between $J[\xi,g_{\alpha\beta}+\delta g_{\alpha\beta}]$
and $J[\xi,g_{\alpha\beta}]$, where $g_{\alpha\beta}+\delta g_{\alpha\beta}$
also solves the
equation of motion, in other words, where $\delta g_{\alpha\beta}$ satisfies
the linearized equations of motion.
We have
\be
\delta J={\cal L}_{\xi}\Theta-i_{\xi}\delta L.
\ee
Utilizing the identity in Eq.(\ref{id1}) and the on-shell condition
$d\Theta=\delta L$, we find
\be
d\Big(\delta Q-i_{\xi}\Theta(g_{\alpha\beta},\delta g_{\alpha\beta})\Big)=0\,,
\ee
where the definition of $\delta$ has been given in previous section.
Evaluating this equation in the region bounded by the horizon and infinity
leads to the first law of thermodynamics
\be
dE=TdS-\int_{\infty}\Theta(g_{\alpha\beta},\delta g_{\alpha\beta}).
\ee
In the context of conformal gravity, the cosmological constant $a_0$ is a parameter of the solution, rather
than of the theory, and hence we may treat $a_0$ as a further thermodynamic variable. Treating the cosmological
constant as a thermodynamic variable has been considered previously. See, for example, \cite{Kastor:2009wy, Cvetic:2010jb,lopapova,Dolan:2013ft}.
Specific to the six-dimensional conformal gravity, by calculating the second term
in the above equation, we
obtain for the general five-parameter solutions
\be
dE=TdS+\Psi_0\, da_0 + \Psi_1 da_1 + \Psi_2 da_2+\Psi_3 da_3\,,
\label{firstlaw}
\ee
where
\bea
&& \Psi_0=
\frac{24\beta V(\Omega_k)}{25}
\Big(50a_0a_5+20a_1a_4-19a_2a_3-6ka_3\Big),\cr
&& \Psi_1=\frac{12\beta V(\Omega_k)}{25}
\Big(3a_2^2+14ka_2-17k^2-5a_1a_3-20a_0a_4\Big),\cr
&& \Psi_2=-\frac{48\beta V(\Omega_k)}{5}a_0a_3,\qquad
\Psi_3=-\frac{72\beta V(\Omega_k)}{5}a_0\Big(a_2-k\Big)\,.
\eea
These quantities satisfy a Smarr-like formula
\be
E=2\Psi_0 a_0+ \Psi_1a_1-  \Psi_3a_3\,,\label{smarr}
\ee
which coincides with the result from dimensional analysis.

   It is straightforward to verify that the explicit three-parameter
black holes given by (\ref{ffn}) and (\ref{arels}) indeed satisfy the first
law (\ref{firstlaw}) and Smarr relation (\ref{smarr}).  It should be emphasised,
however, that the more general five-parameter solutions, which we are only
able to present as infinite series expansions, will also satisfy the first
law and Smarr relation.

\section{Coupling to Conformal Matter}

In four dimensions, Maxwell theory is conformally invariant, and so is
the enlarged system when it is coupled to conformal gravity.
Charged black hole solutions in this theory can be used for studying
some strongly coupled fermionic systems, such as non-Fermi liquids.
In particular, the charged massless Dirac equation can be solved exactly
for a generic frequency $\omega$ and wave number $k$.  Using this, an
explicit expression for the Green function $G(\omega,k)$ was obtained
for general $\omega$ and $k$ in Ref. \cite{Lu:2012ag,Li:2012gh}.
By contrast, such a Green function in the Reissner-Nordstr\o m black hole
geometry can only be obtained explicitly for small $\omega$, and  in the
extremal or near-extremal limit \cite{lmv,flmv}.

The analogous conformal ``matter'' in six dimensions is described by a
2-form potential $B$ whose field strength is $H=dB$.  It is also possible
to write down a conformally-invariant coupling of a
vector potential $A$ coupled quadratically to the Weyl tensor through its
field strength $F=dA$.  Slightly more generally, we may consider the
a conformally-invariant matter Lagrangian of the form
\begin{equation}
{\cal L}_{\rm mat} = \sqrt{-g} \Big( \gamma\, C^{\mu\nu\rho\sigma} F_{\mu\nu} F_{\rho\sigma} - \ft1{12} H^2\Big) + \sigma B\wedge F\wedge F\,,\label{matterlag}
\end{equation}
where the 3-form field strength is now given  by
\begin{equation}
H=dB + \sigma A\wedge F\,,
\end{equation}
with $\sigma$ being a constant.
The Bianchi identity and the equation of motion for $H$ are given by
\begin{equation}
dH=\sigma F\wedge F=d{*H}\,.
\end{equation}
It follows that it is consistent to impose the self-duality condition
$H={*H}$.  Note that the Maxwell field $A$ can be replaced, more
generally, by a
Yang-Mills field without breaking the conformal symmetry.
The rich structure of conformal matter suggests that there should be
a variety of applications of six-dimensional conformal gravity in the
AdS/CFT correspondence

\section{Further Solutions}

  In this section, we present various further solutions of conformal gravity
and of the conformal theories with additional fields that we discussed in
the previous section.  Specifically, section 6.1 contains solutions of
the pure conformal gravity theory, section 6.2 contains solutions of
the conformal theory including a Maxwell field, and sections 6.3
contains solutions in the conformal theory with instead a 2-form potential.

\subsection{Neutral solutions}

\noindent{\bf Lifshitz Black Holes:}

  There are Lifshitz vacuum solutions in the theory described
by (\ref{d6conflag}), given by
\be
ds^2 = -r^{2z} (1+\fft1{r^2}) dt^2 + \fft{\sigma dr^2}{r^2}\,
(1+\fft1{r^2})^{-1} + r^2 d\Omega_4^2\,,
\ee
with $z=0$ or $z=\fft83$:
\bea
z=0:&& \sigma=4\,,\nn\\
z=\fft83:&& \sigma= \fft49\,.
\eea

   We can find explicit black hole solutions that are asymptotic to these
Lifshitz geometries, and that are conformally related to the
Schwarzschild-AdS solution.  For $z=\fft83$, we find the black hole
solution
\be
ds^2 = - r^{16/3}\, f dt^2 + \fft{4 dr^2}{9 r^2 f} + r^2 d\Omega_{4,k}^2\,,
\ee
where
\be
f = r^{-10/3}\, \Big[ -\ft15\Lambda c^2 + k(r^{2/3} + a)^2 -
 m c^{-3} (r^{2/3} + a)^5\Big]\,.
\ee
It is conformally related to the Schwarzschild-AdS metric
\be
d\hat s^2 = -c^2 (k -\fft{m}{\rho^3} -\ft15 \Lambda \rho^2) dt^2 +
\fft{d\rho^2}{(1-\fft{m}{\rho^3} -\ft15\Lambda \rho^2)} +
\rho^2 d\Omega_4^2\,,\label{schwads}
\ee
by $ds^2= \Omega^{-2}\, d\hat s^2$ with $\rho=\Omega r$ and
\be
\Omega =\fft{c}{r^{5/3} + a r}\,.
\ee

For the case $z=0$, we find the black hole
solution
\be
ds^2 = - f dt^2 + \fft{4 dr^2}{ r^2 f} + r^2 d\Omega_{4,k}^2\,,
\ee
where
\be
f = r^2\, \Big[ -\ft15\Lambda c^2 + k(r^{-2} + a)^2 -
 m c^{-3} (r^{-2} + a)^5\Big]\,.
\ee
It is conformally related to the Schwarzschild-AdS metric (\ref{schwads})
by $ds^2= \Omega^{-2}\, d\hat s^2$ with $\rho=\Omega r$ and
\be
\Omega =\fft{c}{r^{-1} + a r}\,.
\ee

\noindent{\bf String solution:}

\bea
ds^2&=&f(r)(-dt^2+dx^2)+f(r)^{-1}(dr^2+r^2d\Omega^2_3),\cr
f &=& \Big(1-\fft{m}{r^2}\Big)^{-\ft12 +\sqrt{\ft32}}\,
   \Big(1+\fft{m}{r^2}\Big)^{-\ft12-\sqrt{\ft32}}\,.
\label{neutralstring}
\eea
This solution has a power-law singularity at $r=\sqrt{m}$.

\subsection{Charged black hole}

The neutral spherically-symmetric black hole solutions (\ref{gs})
can be generalised by turning on the vector field $A$ that enters
the Lagrangian (\ref{matterlag}).  We may consider the
black hole ansatz
\begin{equation}
ds^2 = -f dt^2 + \fft{dr^2}{f} + r^2 d\Omega_{4,k}^2\,,\qquad
A=\phi dt\,,
\end{equation}
where $f$ and $\phi$ are functions of $r$ only.
Letting $\rho=1/r$ and $h(\rho) = r^{-2}f(r)$, we find that the
function $\phi$ satisfies
\begin{equation}
\phi'(\rho) = \fft{q}{h''(\rho)}\,,
\end{equation}
where a derivative is with respect to $\rho$.  The function $h$ then
satisfies
\be
\beta\Big(10hh^{(6)}+30h'h^{(5)}+12h''h^{(4)}-13(h^{(3)})^2-84kh^{(4)}\Big)=15 \gamma (\phi''^2 + \phi' \phi''')\,.
\ee
The general solution that is asymptotic to AdS takes the form
\begin{equation}
f=r^2 \Big(c_0 + \fft{c_1}{r} + \fft{c_2}{r^2} + \fft{c_3}{r^3} +
\fft{c_4}{r^4} + \fft{c_5}{r^5} + \sum_{i=6} \fft{c_i}{r^i}\Big)\,,
\end{equation}
and the coefficients $c_i$ with $i\ge 6$ can be expressed in terms of
the $c_i$ with $i=0,1,\cdots,5$, in a manner analogous to the way
the $b_i$ coefficients in (\ref{gs}) were solved in the neutral case.

We have found two special solutions:
\bigskip

\noindent
{\bf Solution 1}: We find a truncated solution
\begin{equation}
ds^2 = -(c_0 r^2 + c_1 r + c_2) dt^2 + \fft{dr^2}{c_0 r^2 + c_1 r + c_2} + r^2 d\Omega_{4,k}^2\,,\qquad
A= \fft{q}{2c_2\,r} dt\,.
\end{equation}
This solution is analogous to the four-dimensional ``BPS'' black hole
obtained in \cite{Lu:2012am}.  In particular, when the  $c_i$ are chosen
such that $(-g_{tt})$ is a perfect square, the metric has an AdS$_2$
factor in the near-horizon geometry, as in the four-dimensional example.

\noindent
{\bf Solution 2}: If $\beta=0$, we find that the equations can be
solved exactly, giving
\begin{equation}
a=a_0 (1 + \fft{Q}{r})^{3/2}\,,\qquad
f=r^2 \Big(d_0 + \fft{d_1}{r} + \tilde d_0 a\big)\,.
\end{equation}

\subsection{Black dyonic string solutions}

We consider the Lagrangian density of six-dimensional conformal gravity
coupled to a
2-form potential:
\be
e^{-1}{\cal L}=\beta ( 4 I_1 + I_2 - \ft13 I_3  )-\ft1{12}H_{\mu\nu\lambda}H^{\mu\nu\lambda}\,.
\ee
The associated equations of motion are
\be
0=\beta(E^{(1)}_{\mu\nu}-\ft3{25}E^{(2)}_{\mu\nu}-2E^{3}_{\mu\nu}-E^{(4)}_{\mu\nu}+\ft3{10}E^{(5)}_{\mu\nu})
-\ft14(H_{\mu}^{~\lambda\rho}H_{\nu\lambda\rho}-\ft16g_{\mu\nu}H_{\lambda\rho\sigma}H^{\lambda\rho\sigma}).
\ee

\subsubsection{Type I}

\bea
&& ds^2=-f(r)dt^2+r^2dx^2+f(r)^{-1}dr^2+r^2d\Omega_{3,k}^2,\cr
&& H_{(3)}=Qr^{-2}dt\wedge dx\wedge dr+P\Omega_{(3,k)}.
\eea
The solution is given by
\begin{equation}
f=r^2 \Big(a_0 + \fft{a_1}{r} + \fft{a_2}{r^2} + \fft{a_3}{r^3} +
\fft{a_4}{r^4} + \fft{a_5}{r^5}\Big)\,,
\end{equation}
where $a_0$ and $a_1$ can take arbitrary values, $a_3=a_4=a_5=0$ and
\be
\frac{24\beta}{25}(2a_2-k)(a_2+2k)^2-(P^2+Q^2)=0,\quad \frac{24\beta}{25}(a_2+2k)(4a_2^2-a_2k+2k^2)+(P^2+Q^2)=0.
\ee
The above two equations lead to
\bea
&&a_2+2k=0,\qquad (P^2+Q^2)=0,\cr
&&3a_2+k=0,\qquad \ft{40\beta}{9}k^3+(P^2+Q^2)=0,\cr
&&a_2=0,\qquad \ft{96\beta}{25}k^3+(P^2+Q^2)=0.
\eea
Among these solutions, we find that the first one, with vanishing flux,
is actually conformally Einstein. Explicitly,
\be
d\hat{s}^2=\Omega^2ds^2,\quad \Omega^2=r^{-2}\mbox{sech}^2
\frac{\sqrt{k}(x-c)}{\sqrt{2}}\quad \hat{R}_{\mu\nu}=
\frac{5k}{2}\hat{g}_{\mu\nu}\,.
\ee
The conformal metric describes a static soliton located at $x=c$, when $k>0$.

\subsubsection{Type II}
\bea
&& ds^2=\frac{1}{H(r)}(-f(r)dt^2+dx^2)+H(r)(f(r)^{-1}dr^2+r^2d\Omega_{3,k}^2),\cr
&& H_{(3)}=QH(r)^{-2}r^{-3}dt\wedge dx\wedge dr+P\Omega_{(3,k)}.
\eea
One class of solution is found to be
\be
H(r)=1,\quad f(r)=a_0r^2+a_2,\quad \frac{96\beta}{25}(a_2-k)^2(a_2+k)+(P^2+Q^2)=0.
\ee

\subsubsection{Other ans\"atze}

We may consider the following ansatz \cite{Duff:1996cf}:
\bea
&& ds^2=f(r)(-dt^2+dx^2)+f(r)^{-1}(dr^2+r^2d\Omega^2_3),\cr
&& H_{(3)}=Qf^2r^{-3}dt\wedge dx\wedge dr+P\Omega_{(3)}\,,\label{otherstring}
\eea
where $\Omega_{(3)}$ is the volume form of the unit $S^3$.
Adopting the above ansatz, the equations of motion for the 2-form potential
are satisfied automatically. Before presenting the equation following
from the metric variations, it is useful to make the field redefinitions
and coordinate transformation
\be
r=e^{\rho}\,,\quad f(r)=e^{h(\rho)+\rho}\,,\quad \dot{h}(\rho)=W(\rho)\,,
\ee
where a dot denotes a derivative with respect to $\rho$.
In terms of $W(\rho)$ we find
\bea
0&=&8-8W^2-8W^4+8W^6-24W^2\dot{W}-136W^4\dot{W}+6\dot{W}^2+14W^2\dot{W}^2-16\dot{W}^3-12W\ddot{W}\cr
&&-28W^3\ddot{W}+48W\dot{W}\ddot{W}+5\ddot{W}^2+40W^2\dddot{W}-
 10\dot{W}\dddot{W}+10W\ddddot{W}+\frac{25}{12\beta}(P^2+Q^2)\,.
\eea
A class of solutions of this equation is given by
\be
f(r)=r^a,\qquad \ft{96}{25}\beta(a-2)^2a^2(a^2-2a+2)+(P^2+Q^2)=0\,.
\ee
For this solution to be real $\beta$ must be negative, coinciding with
condition under which the energy and entropy of the AdS black holes are
positive. Especially, when $a=2$, the solutions is AdS$_3\times S_3$ with
vanishing string charges. By a conformal scaling, the solutions can be
mapped to
\be
d\hat{s}^2=r^{-2a}ds^2=(-dt^2+dx^2)+d\rho^2+(a-1)^2\rho^2d\Omega_3^2\,.
\ee
This has a conical singularity at the origin of the transverse space of
the string.

   To obtain AdS$_3\times S_3$ solutions with non-trivial flux, we
reparametrize the metric and $H_{(3)}$ as
\bea
&&ds^2=r^2(-dt^2+dx^2)+r^{-2}dr^2+a^2d\Omega_3^2,\cr%
&& H_{(3)}=Qf^2r^{-3}dt\wedge dx\wedge dr+Pa^3\Omega_{(3)}\,.
\eea
The equations of motion are solved provided that
\be
\ft{96}{25}\beta(a^2-1)^2(a^2+1)+a^6(P^2+Q^2)=0\,.
\ee

  The theory also admits AdS$_3\times \tilde{S}_3$ as a solution,
where $\tilde{S}_3$ is a squashed 3-sphere.
The AdS$_3\times \tilde{S}_3$ metric is given by
\be
 ds^2=r^2(-dt^2+dx^2)+r^{-2}dr^2+a^2\sigma_3^2+(\sigma_1^2+\sigma_2^2)\,.
\ee
If we choose the vielbeins to be
\be
e^0=rdt,\quad e^1=rdx\,,\quad e^2=r^{-1}dr\,,\quad e^3=a\sigma_3\,,
\quad e^4=\sigma_2\,,\quad e^5=\sigma_1\,,
\ee
and $H_{(3)}$ to be given by
\be
 H_{(3)}=Qe^0\wedge e^1 \wedge e^2+Pe^3\wedge e^4 \wedge e^5\,,
\ee
then we obtain a solution when
\be
\ft{96}{25}\beta(a^2-1)(41-63a^2)+(P^2+Q^2)=0\,.
\ee

\section{AMD Charge for General Cubic Curvature Theories}

   In this section we apply the conformal methods developed
by Ashtekar, Magnon and Das (AMD) \cite{Ashtekar:1984zz,ashdas} for calculating
conserved charges in asymptotically AdS backgrounds to the case of
cubic-curvature theories in arbitrary dimensions.
The AMD conserved quantities are extracted from the leading fall-off
of the electric part of the Weyl tensor. The fall-off rate of the
curvature is weighted by a smooth function $\Omega$, with the conformal
boundary ${\cal I}$ being defined at $\Omega=0$.  For further details
about the conditions on the choice of $\Omega$,the reader is referred to
Refs.\
\cite{Ashtekar:1984zz, ashdas}; here we only mention some
necessary points. For a $d$-dimensional asymptotically AdS spacetime
($d\geq4$), on the boundary ${\cal I}$ we require
\begin{eqnarray}
 & &\hat{g}_{\mu\nu} =\Omega^2 g_{\mu\nu}, \\
  &&\hbox{At}\ \Omega= 0,\quad \n_\mu=\del_\mu\Omega\neq0,\\
    & &\n_\mu\n^\mu=\frac{1}{\ell^2},\quad\hat{\nabla}_\mu\n_\nu=0\,.
\end{eqnarray}
Since ${\cal I}$ is defined to be at $\Omega=0$, it follows that
$n_{\mu}$ is  normal vector on the boundary ${\cal I}$.
Near the boundary,
\begin{eqnarray}
  &&R_{\mu\nu\lambda\rho}\rightarrow-\frac{1}{\ell^2}(g_{\mu\lambda}g_{\nu\rho}-g_{\mu\rho}g_{\nu\lambda}),\label{asRiem}\\
&&  T_{\mu\nu}\rightarrow\Omega^{d-2}\tau_{\mu\nu},\label{asT}\\
   &&C_{\mu\nu\lambda\rho}\rightarrow\Omega^{d-5}
 K_{\mu\nu\lambda\rho}\label{asWeyl}\,,
\end{eqnarray}
where $g_{\mu\nu}$ is the physical metric, the hatted quantities  are
referred to the conformal metric $\hat{g}_{\mu\nu}$, and $T_{\mu\nu}$ is the
energy momentum tensor. As first noticed in
\cite{Okuyama:2005fg}, the condition (\ref{asRiem}) is
required in higher-curvature theories in order to ensure that the metric
that satisfies the  equations of motion is indeed asymptotically
AdS. (In Einstein gravity, by contrast, as discussed in
\cite{Ashtekar:1984zz,ashdas}, Eq. (\ref{asRiem})
is {\it implied} by the Einstein equations together with
Eqs. (\ref{asT}) and (\ref{asWeyl}).)

 For any theory of gravity with the equations
\begin{equation}\label{}
    E_{\mu\nu}=8\pi G_{(d)}T_{\mu\nu}\,,
 \end{equation}
one can show that
\begin{equation}\label{}
    \Omega^{-(d-3)}(\nabla_{[\lambda}P_{\mu]\nu})\n^{\lambda}\n^{\nu}\xi^{\mu}=\frac{d-2}{2\ell^2}8\pi
    G_{(d)}\tau_{\mu\nu}\n^{\nu}\xi^{\mu}+O(\Omega)\,,
\end{equation}
where
\begin{equation}\label{}
    P_{\mu\nu}\equiv E_{\mu\nu}-\ft{1}{d-1}g_{\mu\nu}
 E_{\lambda\rho}g^{\lambda\rho}\,.
\end{equation}
In general, the leading fall-off of
$(\nabla_{[\lambda}P_{\mu]\nu})\n^{\lambda}\n^{\nu}\xi^{\mu}$ is of the
order $\Omega^{d-3}$,
and can be expressed as
$-\ft{(d-2)}{2(d-3)}\Xi\hat{\nabla}^\rho(K_{\lambda\mu\nu\rho}\n^{\lambda}\n^{\nu}\xi^{\mu})\Omega^{d-3}$.
The conserved quantity associated to the Killing vector $\xi^\mu$ can
be defined, when $\tau_{\mu\nu}$ vanishes on the boundary, as
\begin{equation}\label{}
Q_{\xi}[C]=-\frac{\ell\Xi}{8\pi G_{(d)}(d-3)}\int_C
d\hat{S}_{(d-2)}^{m} \hat{\cal E}_{ m n}\xi^n\,,\qquad \hat{\cal
E}_{mn}\equiv \ell^2K_{\lambda m\rho n}\n^{\lambda}\n^{\rho}\,,\label{amd}
\end{equation}
where the indices $m$ and $n$ label the coordinates on the $(d-1)$-dimensional
boundary ${\cal I}$, since the electric part
of the Weyl tensor, $\hat{\cal E}_{ m n}$, has no components in the
normal direction. $C$ is a $(d-2)$ dimensional spherical cross-section on
${\cal I}$.

Consider the Lagrangian for the general class of cubic-curvature theories
of the form
\begin{eqnarray}
  16\pi G_{(d)}e^{-1}{\cal L}&=& - 2 \Lambda + R + \alpha_1 {\cal L}_{\rm
GB} + \alpha_2 R^2 + \alpha_3 R_{\mu\nu} R^{\mu\nu}+\beta_1RR_{\mu\nu} R^{\mu\nu}+ \beta_2R^3\nn\\
     &&+ \beta_3R_{\mu\lambda\nu\rho}R^{\mu\nu}R^{\lambda\rho}+\beta_4R_{\mu\nu}\Box R^{\mu\nu}+
     \beta_5R\Box R+\beta_6R_{\mu}^{~\nu}R_{\nu}^{~\lambda}R_{\lambda}^{~\mu}\nn \\
     &&+\beta_7R_{\mu\nu}R^{\mu\lambda\rho\sigma}R^{\nu}_{~\lambda\rho\sigma}
    +\beta_8RR^{\mu\nu\lambda\rho}R_{\mu\nu\lambda\rho} + \beta_9R^{\mu\nu}_{~~\lambda\rho}R^{\lambda\rho}_{~~\sigma\delta}R^{\sigma\delta}_{~~\mu\nu}
    \nn \\
     &&  +\beta_{10}R^{\mu~\nu}_{~\lambda~\rho}
R^{\lambda~\rho}_{~\sigma~\delta}R^{\sigma~\delta}_{~\mu~\nu}\label{cclag}\,.
\end{eqnarray}
The AMD formula for quadratic-curvature theories has been obtained
in \cite{Pang:2011cs}.  By repeating the procedure (the corrections to
the equations of motion from cubic curvature terms are presented in Appendix),
we can obtain the contributions from the cubic-curvature terms to
the AMD charges. In the general cubic-curvature theories Eq.(\ref{cclag}),
the AMD charges take the same form as in Eq.(\ref{amd}), with
the coefficient of proportionality $\Xi$ given by
\bea
\Xi&=& 1 + R_0 \left[ 2 \alpha_1 \frac{(d-3)(d-4)}{d ( d - 1 )} + 2
\alpha_2 + \frac{2\alpha_3 }{d } \right]+R^2_0\bigg[\frac{3\beta_1}{d}+3\beta_2+\frac{3\beta_3}{d^2}+\frac{3\beta_6}{d^2}\nn\\
&&+\beta_7\frac{2(9-2d)}{d^2(d-1)}+\beta_{8}\frac{2(9-2d)}{d(d-1)}+\beta_9\frac{12(7-2d)}{d^2(d-1)^2}
+\beta_{10}\frac{3(3d-8)}{d^2(d-1)^2}\bigg],\nn\\
R_0&=&-\frac{d(d-1)}{\ell^2}\label{camd}\,.
\eea
We notice that the terms $R_{\mu\nu}\Box R^{\mu\nu}$ and $R\Box R$ do not
contribute to the charge, for solutions whose asymptotic behavior obeys
Eqs.(\ref{asRiem}) and (\ref{asWeyl}).

  As a check of the above formula, we can calculate the charge for the case
of the  six-dimensional Euler density
\be
E_6=\ft18\epsilon_{\mu_1\nu_1 \mu_2\nu_2 \mu_3\nu_3} \epsilon^{\rho_1
\sigma_1 \rho_2 \sigma_2 \rho_3\sigma_3} R^{\mu_1
\nu_1}{}_{\rho_1\sigma_1} R^{\mu_2 \nu_2}{}_{\rho_2\sigma_2}
R^{\mu_3 \nu_3}{}_{\rho_3\sigma_3}\,.
\ee
In terms of the quantities in Eq.(\ref{cclag}), the Euler density $E_6$
corresponds to the combination of cubic-curvature terms with coefficients
\bea
&&\beta_1=-12,\quad\beta_2=1,\quad\beta_3=24,\quad\beta_4=0,\quad\beta_5=0,\nn\\
&&\beta_6=16,\quad\beta_7=-24,\quad\beta_8=3,\quad\beta_9=4,\quad\beta_{10}
=-8\,.
\eea
Inserting these coefficients into Eq.(\ref{camd}), we find that
the coefficient $\Xi$ for $E_6$ is given by
\be
\Xi_{E_6}=R^2_0\frac{24(d-3)(d-4)(d-5)(d-6)}{d^2(d-1)^2}\,,
\ee
which indeed vanishes for $d=6$ as expected.

  The AMD formula for general cubic-curvature theories can also be used
for finding the criticality condition and computing the conserved
quantities in quasi-topological gravity \cite{Oliva:2010eb,Myers:2010ru}.
The $d$-dimensional quasi-topological gravity is defined by the action
\bea
I&=&\frac{1}{16\pi G_d}\int d^dx\biggl(\frac{(d-1)(d-2)}{L^2}+R+\frac{\lambda L^2}{(d-3)(d-4)}{\cal X}_4\cr
&&\qquad\quad-\frac{8(2d-3)}{(d-6)(d-3)(3d^2-15d+16)}\mu
 L^4{\cal Z}_D\biggr)\,,
\eea
where ${\cal X}_4$ is the Gauss-Bonnet combination and ${\cal Z}_D$ is
the quasi-topological combination consisting of cubic-curvature terms with
\bea
&&\beta_1=\frac{-3(3d - 4)}{2(2d - 3)(d - 4)},\quad\beta_2=\frac{3d}{8(2d- 3)(d - 4)},\quad\beta_3=\frac{3d}{(2d - 3)(d - 4))},\quad\beta_4=0,\nn\\ &&\beta_5=0,\quad
\beta_6=\frac{6(d - 2)}{(2d - 3)(d - 4)},\quad
\beta_7=-\frac{3(d - 2)}{(2d - 3)(d - 4)},\quad
\beta_8=\frac{3(3d - 8)}{8(2d - 3)(d - 4)},\nn\\
&&\beta_9=0,\quad\beta_{10}=1\,.
\eea
This  theory has as a solution the asymptotically AdS metric
\be
ds^2=-(k+\frac{r^2}{L^2}f)dt^2+\frac{dr^2}{k+\frac{r^2}{L^2}f}+
r^2 d\Omega_k^2\,,
\ee
where $f(r)$ satisfies the cubic equation
\be
(1-\frac{\omega^{d-1}}{r^{d-1}})-f+\lambda f^2+\mu f^3=0\,.
\ee
To compute the mass of these black holes using the AMD formula
Eq.(\ref{camd}), we choose $\Omega=1/r$.  We then find that
\bea
&&C_{t\Omega t\Omega}\rightarrow-\ft12(d-2)(d-3)\Omega^{d-5}\frac{\omega^{d-1}}{L^2(1-2\lambda f_{\infty}-3\mu f_{\infty})}\cr
&&\Xi=1-2\lambda f_{\infty}-3\mu f_{\infty}\,,
\eea
where $f_\infty$ denotes the asymptotic value of $f(r)$ as $r$ tends to
infinity.
Therefore Eq.(\ref{amd}) gives the mass of the black holes in
quasi-topological gravity as
\be
M=\frac{(d-2)\omega^{d-1}V(\Omega_k)}{16\pi G_{(d)}L^2}\,.
\ee
The temperature and entropy (using Wald's formula) are
\be
T=\frac{(d-1)\omega}{4\pi L^2}\,,\qquad
S=\frac{\omega^{d-2}V(\Omega_k)}{4G_{(d)}}+\epsilon_k\,,
\ee
where $\epsilon_k$ is a certain constant depending on the topology of horizon.
It is straightforward to check that the first law of thermodynamics
holds in this case.

\section{Tricritical Gravity in Six Dimensions }

  ``Critical gravity'' is the name given to higher-derivative theories of
gravity that admit AdS backgrounds, and which generically describe
massive as well as massless spin-2 modes, in the special case where the
parameters of the theory are tuned such that the massive spin-2 modes become
massless.  One example is the chiral point in three-dimensional
topologically-massive gravity, discussed in
\cite{lisost}, and another is the four-dimensional critical theory
discussed in \cite{lpcritical}, where a Weyl-squared term with an
appropriately tuned coefficient is added to cosmological Einstein gravity.
In that case, with a fourth-order Lagrangian, there is one massive spin-2
excitation in addition to the usual massless spin-2.  In theories of
the kind we are considering in this paper, with sixth-order Lagrangians,
there are in general two massive spin-2 excitations in addition to the
massless spin-2, and so the possibility of tuning the parameters so that
all three are massless arises.  This is known as tricritical gravity.

\subsection{The theory}

In six dimensions, one tricritical gravity model has been constructed
in which the scalar modes do not propagate \cite{lupapo}. The Lagrangian
for this theory is given by
\be
16\pi G_{(6)}\sigma^{-1}e^{-1}{\cal L}_6=- 2 \Lambda + R+\ft12\alpha
C^{\mu\nu\lambda\rho}C_{\mu\nu\lambda\rho}-{\cal L}_{\rm conf}\,,
\ee
where ${\cal L}_{\rm conf}$ is defined in Eq.(\ref{d6conflag}) and
$\sigma$ is the overall sign. In the AdS$_6$ background, the spin-2
modes satisfy
\be
-(\Box+2)\biggl(1+\ft32\alpha(\Box+6)+\beta(\Box+6)(\Box+8)\biggr)
h_{\mu\nu}=0\,.
\ee
The tricriticality condition is achieved when
\be
\alpha=-\frac{5}{12},\qquad\beta=\frac{1}{16}\,,
\ee
where the AdS ``radius'' has been set to 1.

   Another tricritical model has the Lagrangian
\be
16\pi G_{(6)}\sigma^{-1}e^{-1}{\cal L}_6=- 2 \Lambda +
R+\ft14\alpha(R^{\mu\nu}R_{\mu\nu}-\ft3{10}R^2)-{\cal L}_{\rm conf}\,.
\ee
This also admits all Einstein metrics as solutions.
 The spin-2 modes in this case satisfy
\be
-(\Box+2)\biggl(1+\ft14\alpha(\Box+10)+\beta(\Box+6)(\Box+8)\biggr)
h_{\mu\nu}=0\,,
\ee
and the tricriticality condition is achieved when
\be
\alpha=-\frac{5}{7},\qquad\beta=\frac{1}{56}\,.
\ee

    In both models, at the tricritical point the spin-2 modes satisfy
\be
-(\Box+2)^3h_{\mu\nu}=0\label{tceom}\,.
\ee
Massless, massive and log modes of the spin-2 field $h_{\mu\nu}$ in AdS$_6$ were obtained in \cite{Chen:2011in}.

\subsection{Consistent boundary conditions in tricritical gravity}

Our starting point is the AdS$_6$ metric coordinatized as
\be
ds^2 = \ell^2\biggl(-\cosh^2\rho dt^2 + d\rho^2 + \sinh^2\rho (d\theta_1^2
+ \sin^2\theta_1
(d\theta^2_2+ \sin^2\theta_2d\phi_1^2
+ \cos^2\theta_2d\phi_2^2))\biggr)\,.\label{AdS6}
\ee
Near the AdS$_6$ boundary, the solutions to Eq.(\ref{tceom})
\cite{Chen:2011in} have the fall off behavior
\be
h_{55}={\cal O}(\rho^2e^{-7\rho})\,,\qquad h_{5i} = {\cal O}(\rho^2e^{-5\rho})
\,,\qquad h_{ij}= {\cal O}(\rho^2e^{-3\rho})\,,\label{lmodes}
\ee
where $x^i=\{t,\theta_1,\theta_2,\phi_1,\phi_2\}$ for $0\le i\le 4$ and
$x^5=\rho$.
This implies that near the AdS$_6$ boundary, the Weyl tensor has the
asymptotic behavior
\be
C_{\mu\nu\lambda\rho}\rightarrow\Omega\,\biggl(
\log^2\Omega \, J_{\mu\nu\lambda\rho}
+\log\Omega \, L_{\mu\nu\lambda\rho}+K_{\mu\nu\lambda\rho}\biggr)\,,
\ee
where $\Omega$ approaches $e^{-2\rho}$ near the boundary. In this case,
we find that
for the two tricritical models,
$(\nabla_{[\lambda}P_{\mu]\nu})\n^{\lambda}\n^{\nu}\xi^{\mu}$ remains
of order $\Omega$, and therefore one can obtain finite AMD charges for
these two models.

   Explicitly, for the first tricritical model, the AMD charge at the
tricritical point is given by
\be
Q_{\xi}[C]_1=-\frac{25\ell\sigma}{192\pi G_{(6)}}\int_C
d\hat{S}_{4}^{m} \hat{\cal E}_{ m n}\xi^n,~~~ \hat{\cal
E}_{mn}\equiv \ell^2J_{\lambda m\rho n}\n^{\lambda}\n^{\rho}\,.
\ee

   Similarly, for the second tricritical model, the AMD charge at the
tricritical point is given by
\be
Q_{\xi}[C]_2=-\frac{25\ell\sigma}{672\pi G_{(6)}}\int_C
d\hat{S}_{4}^{m} \hat{\cal E}_{ m n}\xi^n,~~~ \hat{\cal
E}_{mn}\equiv \ell^2J_{\lambda m\rho n}\n^{\lambda}\n^{\rho}\,.\label{amd2}
\ee

Asymptotic Killing vectors should be compatible with the boundary
conditions (\ref{lmodes}),, implying that they should obey
\be
{\cal L}_\xi \, g_{55}={\cal O}(\rho^2e^{-7\rho})\,,\qquad
{\cal L}_\xi \, g_{5i} = {\cal O}(\rho^2e^{-5\rho})
\,,\qquad {\cal L}_\xi\, g_{ij}= {\cal O}(\rho^2e^{-3\rho})\,.
\ee
Vector fields $\xi$ satisfying these equations
(modulo ``trivial'' diffeomorphisms) generate the
asymptotic symmetry group. We denote the Killing vector fields
by $U_{ab}^{\mu}$ $(a,b=1,\ldots,7)$. Since in the
coordinate system used in Eq.(\ref{AdS6}), these obey
$U_{ab}^{\mu}={\cal O}(1)$, we find that the asymptotic Killing vector
fields can only differ from the Killing vectors at the order
\be
\xi^{\mu}=\ft12\xi^{ab}_{\infty}U_{ab}+{\cal O}(\rho^2e^{-7\rho})\,,
\ee
where $\xi^{ab}_{\infty}$ is constant. The boundary conditions
Eq.(\ref{lmodes}) can be verified to be consistent, yielding well-defined
charges that are finite, integrable and conserved. It can be shown that
the associated asymptotic symmetry group is still $SO(2,5)$.

\section{Conclusions}

   In this paper, we have studied some aspects of conformally-invariant
gravities in six dimensions.  Unlike in four dimensions where there is a
unique theory that is polynomial in the curvature or its derivatives
(described by a Weyl-squared Lagrangian), in six dimensions
there are three such independent conformally-invariant terms that could be
considered.  However, if we impose the additional requirement that, like
in four dimensions, Einstein metrics should also be solutions of the theory,
then this implies that a unique linear combination of the three terms
is singled out. It is this specific theory that has formed the focus of most
of our attention in this paper, since it has the advantage that at least
some solutions, namely Einstein metrics and their conformal scalings,
can be obtained explicitly.

   Using the freedom to perform coordinate transformations and conformal
scalings, the general ansatz for spherically-symmetric black holes can be
expressed in terms of a single function of the the radius.  This function
obeys a 5th-order differential equation which, unfortunately, we have not
been able to solve in closed form in general.  We were, however, able to
construct the general solution as an infinite series expansion, characterised
by the expected number of five independent parameters.  Within this class
of solutions is a three-parameter subset for which the series expansion
terminates.  This closed-form class of solutions corresponds precisely
to the standard Schwarzschild-AdS metrics, and their spherically-symmetric
conformal scalings.  We studied the thermodynamics of the black holes,
obtaining a first law for the five-parameter family of solutions, and
verifying that this was indeed satisfied by the explicit closed-form
subset of solutions.

We considered also some more general conformal theories in six dimensions,
in which conformally-invariant ``matter'' is coupled to conformal gravity.
Specifically, we looked at a bilinear coupling of a Maxwell field strength
to the Weyl tensor, and also kinetic and Chern-Simons terms involving
a 2-form potential.  We obtained a variety of further solutions for these
theories, and also for the pure conformal gravity.

In our work, we concentrated on the particular choice of six-dimensional
conformal gravity for which conformally-Einstein metrics are also solutions.
It would be of interest also to study the broader class of conformal
gravities in six dimensions for which this is no longer the case. It may
not be easy, within the broader class of theories, to obtain explicit
closed-form solutions, but nevertheless it could be of interest to
investigate black hole solutions, and their thermodynamics.

A further interesting question is whether any of the six-dimensional conformal
gravities could be supersymmetrised.  As far as we are aware, there are
no known obstacles to doing this, other than the complexity of the
problem.  If it could be achieved, then it would presumably be an
off-shell theory, since experience suggests that this is the only way
in which one is likely to be able to construct a higher-derivative
supergravity that does not require an infinity of higher-order terms
(such as in string theory).

\bigskip

\section*{Acknowledgements}

We are grateful to Ergin Sezgin for useful conversations.
The research of H.L. is supported in part by
NSFC grants 11175269 and 11235003.
Y.P. and C.N.P. are supported in part by DOE grant DE-FG03-95ER40917.

\bigskip

\appendix

\section{Necessary Condition for Conformally Einstein}

 In this appendix, we present a detailed derivation of a necessary condition
for a $d$-dimensional metric to be conformally Einstein.  This condition
was derived first in four dimensions in \cite{koneto}, and
subsequently, in arbitrary dimensions, in \cite{gonu}.

  A $d$-dimensional spacetime with metric $g_{ab}$ is conformally Einstein
if there exists a conformal transformation to a new metric
$\hat{g}_{ab}=\Omega^2g_{ab}$ such that
\be
\hat{R}_{ab}-\ft1d\hat{g}_{ab}\hat{R}=0\,,
\ee
or, equivalently,
\be
\hat{P}_{ab}-\ft1d\hat{g}_{ab}\hat{P}=0\,,\label{hatP}
\ee
where
\be
P_{ab} \equiv -
\fft{1}{
(d - 2)}\, R_{ab} +\fft{
1}
{2(d - 1)(d - 2)}\, R g_{ab}\,.
\ee

Defining $\Upsilon_a\equiv\nabla_a\ln\Omega$, then from the conformal
transformation of the Ricci scalar and Ricci tensor we have
\be
\Omega^{-2}(R+2(d-1)\nabla^{c}\Upsilon_c+(d-1)(d-2)\Upsilon_c
\Upsilon^c)= \hbox{constant}\,,\label{eqxx}
\ee
\be
R_{ab}-\fft1d\, g_{ab}R+(d-2)\nabla_a\Upsilon_b-
\fft{d-2}{d}g_{ab}\, \nabla_c\Upsilon^{c}-(d-2)\Upsilon_a\Upsilon_b
+\fft{d-2}{d}g_{ab}\, \Upsilon^c\Upsilon_c=0\,,
\ee
and (\ref{hatP}) becomes
\be
P_{ab}-\ft1dg_{ab}P-\nabla_a\Upsilon_b+\fft1d\,
g_{ab}\nabla_c\Upsilon^c+\Upsilon_a\Upsilon_b
-\fft1d\, g_{ab}\Upsilon^c\Upsilon_c=0\,.
\ee
Taking a derivative of (\ref{eqxx}) gives
\bea
0&=&\nabla_aR-2R\Upsilon_a-4(d-1)\Upsilon_a\nabla^c
\Upsilon_c-2(d-1)(d-2)\Upsilon_a\Upsilon^c\Upsilon_c\nn\\
&&+2(d-1)\nabla_a\nabla_c\Upsilon^c+2(d-1)(d-2\Upsilon^c\nabla_a
\Upsilon_c\nn\\
&=&\nabla_aP-2P\Upsilon_a+2\Upsilon_a\nabla^c\Upsilon_c+(d-2)
\Upsilon_a\Upsilon^c\Upsilon_c
-\nabla_a\nabla_c\Upsilon^c-(d-2)\Upsilon^c\nabla_a\Upsilon_c\,.
\eea
Using this, we obtain
\be
\nabla_{[a}P_{b]c}+\ft12C_{abcd}\Upsilon^{d}=0\,.
\ee
Using
\be
\nabla^dC_{abcd} = 2(d - 3)\nabla_{[a}P_{b]c}\,,
\ee
we finally obtain
\be
\nabla^dC_{abcd}+(d-3)\Upsilon^dC_{abcd}=0\,.
\ee
This necessary condition must be satisfied by any conformally Einstein metric.

\section{Equations of General Cubic Curvature}

In this appendix, we present the detailed results for the variations of
each of the terms in the Lagrangian (\ref{cclag}).  In particular, this
includes the results needed for obtaining the equations of motion
(\ref{d6eom}) for the conformally-invariant theory that forms the focus
of most of our attention in this paper.

\begin{eqnarray}
1):&&RR^{\mu\nu}R_{\mu\nu}\Rightarrow \cr 
&&E^{(1)}_{\mu\nu}=
 (\Box (R_{\lambda\sigma}R^{\lambda\sigma})+\nabla_{\lambda}\nabla_{\sigma}
(RR^{\lambda\sigma}) -\ft{1}{2}RR_{\lambda\sigma}R^{\lambda\sigma})g_{\mu\nu}+R_{\lambda\sigma}R^{\lambda\sigma}R_{\mu\nu} +
2RR_{\lambda\mu} R^{\lambda}_{~\nu}\cr 
&&\qquad\quad+\Box(RR_{\mu\nu})-\nabla_{\mu}\nabla_{\nu}(R_{\lambda\sigma}R^{\lambda\sigma}) -
\nabla_{\lambda}\nabla_{\mu}(RR^{\lambda}_{~\nu})
-\nabla_{\lambda}\nabla_{\nu}(RR^{\lambda}_{~\mu})\,,\cr 
2):&&R^3\Rightarrow\cr 
&&E^{(2)}_{\mu\nu}=(3\Box R^2-\ft{1}{2}R^3)g_{\mu\nu}+3R^2R_{\mu\nu} -
3\nabla_{\mu}\nabla_{\nu}R^2\,,\cr 
3):&&R^{\mu\nu}R^{\lambda\rho}R_{\mu\lambda\nu\rho}
\Rightarrow\cr 
&&E^{(3)}_{\mu\nu}=-\ft{1}{2}R^{\sigma\delta}R^{\lambda\rho}
R_{\sigma\lambda\delta\rho}g_{\mu\nu} + \ft{3}{2}R^{\rho\sigma}
R_{\rho\mu\sigma\lambda} R^{\lambda}_{~\nu}
+\ft{3}{2}R^{\rho\sigma} R_{\rho\nu\sigma\lambda}
R^{\lambda}_{~\mu}\cr 
&&\qquad\quad +\Box(R^{\rho\sigma}R_{\rho\mu\sigma\nu}) +
\nabla^{\sigma}\nabla^{\delta} (R^{\lambda\rho}
R_{\lambda\sigma\rho\delta})g_{\mu\nu}\cr 
&&\qquad\quad -\nabla^{\lambda}\nabla_{\mu} (R^{\rho\sigma}
R_{\rho\lambda\sigma\nu})-\nabla^{\lambda}\nabla_{\nu} (R^{\rho\sigma}
R_{\rho\lambda\sigma\mu})\cr 
&&\qquad\quad -\nabla_{(\sigma}\nabla_{\lambda)}
(R_{\mu}^{~\sigma}R_{\nu}^{~\lambda}) +
\nabla_{\sigma}\nabla_{\lambda}(R_{\mu\nu}R^{\sigma\lambda}),\cr 
4):&&R^{\mu\nu}\Box R_{\mu\nu} = -g^{\mu\nu}
\nabla_{\mu} R^{\lambda\rho}\nabla_{\nu}R_{\lambda\rho}
\Rightarrow\cr 
&&E^{(4)}_{\mu\nu}= \ft{1}{2} g_{\mu\nu}(g^{\sigma\delta} \nabla_{\sigma}
R^{\lambda\rho} \nabla_{\delta}R_{\lambda\rho}) -(2\nabla^{\sigma}
R_{\mu\lambda}\nabla_{\sigma} R_{\nu}^{~\lambda} +
\nabla_{\mu}R_{\sigma\lambda} \nabla_{\nu}R^{\sigma\lambda})\cr 
&&\qquad\quad +2\nabla_{\lambda} (R_{\sigma(\mu}\nabla_{\nu)}
R^{\lambda\sigma}) +2\nabla_{\lambda} (\nabla^{\lambda}
R^{\sigma}_{~(\nu}R_{\mu)\sigma}) -2\nabla_{\sigma}(\nabla_{(\mu}
R_{\nu)\lambda}R^{\sigma\lambda})\cr 
&&\qquad\quad + \Box^2R_{\mu\nu}+\nabla_{\sigma}\nabla_{\lambda}\Box
R^{\sigma\lambda}g_{\mu\nu}-\nabla_{\lambda}\nabla_{\nu}(\Box
R^{\lambda}_{~\mu})-\nabla_{\lambda}\nabla_{\mu}(\Box
R^{\lambda}_{~\nu}),\cr 
5):&&R\Box R=-g^{\mu\nu}\nabla_{\mu}R\nabla_{\nu}R
\Rightarrow \cr 
&&E^{(5)}_{\mu\nu}= \ft{1}{2}g_{\mu\nu}
(g^{\sigma\lambda}\nabla_{\sigma}R\nabla_{\lambda}R)
-\nabla_{\mu}R\nabla_{\nu}R+2(\Box R)R_{\mu\nu}\cr 
&&\qquad\quad +2(\Box^2R)g_{\mu\nu}-2\nabla_{\mu}\nabla_{\nu}\Box R,\cr 
6)&&:R_{\mu}^{~\nu}R_{\nu}^{~\lambda}R_{\lambda}^{~\mu}
\Rightarrow \cr
&&E^{(6)}_{\mu\nu}=-\ft12g_{\mu\nu}R_{\lambda}^{~\rho}R_{\rho}^{~\sigma}R_{\sigma}^{~\lambda}
+3R_{\lambda\mu}R_{\rho\nu}R^{\lambda\rho}+\ft32\Box(R_{\mu}^{~\lambda}R_{\lambda\nu})
+\ft32\nabla_{\rho}\nabla_{\sigma}(R^{\rho}_{~\lambda}R^{\lambda\sigma})g_{\mu\nu} \cr
&&\qquad\quad-\ft32\nabla_{\lambda}\nabla_{\nu}(R^{\lambda}_{~\rho}R^{\rho\mu})
-\ft32\nabla_{\lambda}\nabla_{\mu}(R^{\lambda}_{~\rho}R^{\rho\nu}),\cr
7)&&:R_{\mu\nu}R^{\mu\lambda\rho\sigma}R^{\nu}_{~\lambda\rho\sigma}
\Rightarrow \cr
&&E^{(7)}_{\mu\nu}=-\ft12g_{\mu\nu}R_{\delta\tau}R^{\delta\lambda\rho\sigma}R^{\tau}_{~\lambda\rho\sigma}
+\ft12g_{\mu\nu}\nabla_{\delta}\nabla_{\tau}(R^{\delta\lambda\rho\sigma}R^{\tau}_{~\lambda\rho\sigma})
+\ft12\Box(R_{\mu}^{~\lambda\rho\sigma}R_{\nu\lambda\rho\sigma})\cr
&&\qquad\quad-\ft12\nabla_{\delta}\nabla_{\mu}(R^{\delta\lambda\rho\sigma}R_{\nu\lambda\rho\sigma})
-\ft12\nabla_{\delta}\nabla_{\nu}(R^{\delta\lambda\rho\sigma}R_{\mu\lambda\rho\sigma})
+R_{\delta\sigma}R_{~\mu}^{\delta~\lambda\rho}R^{\sigma}_{~\nu\lambda\rho}\cr
&&\qquad\quad+2R_{\delta\sigma}R_{~~~\mu}^{\delta\lambda\rho}R^{\sigma}_{~\lambda\rho\nu}
-2\nabla_{\rho}\nabla_{\sigma}(R_{\lambda(\mu}R_{\nu)}^{~\lambda\rho\sigma})
-2\nabla_{\rho}\nabla_{\sigma}(R_{\lambda(\mu}R_{~~~~\nu)}^{\lambda\rho\sigma})\cr
&&\qquad\quad +2\nabla_{\rho}\nabla_{\sigma}(R^{\rho}_{~\lambda}R^{\lambda~~\sigma}_{~(\mu~\nu)}),\cr
8)&&:RR^{\mu\nu\lambda\rho}R_{\mu\nu\lambda\rho}
\Rightarrow\cr
&&E^{(8)}_{\mu\nu}=-\ft12g_{\mu\nu}RR^{\lambda\rho\sigma\delta}R_{\lambda\rho\sigma\delta}
+R_{\mu\nu}R^{\lambda\rho\sigma\delta}R_{\lambda\rho\sigma\delta}
+2RR_{\mu\lambda\rho\sigma}R_{\nu}^{~\lambda\rho\sigma}
+g_{\mu\nu}\Box(R^{\lambda\rho\sigma\delta}R_{\lambda\rho\sigma\delta})\cr
&&\qquad\quad-\nabla_{\mu}\nabla_{\nu}(R^{\lambda\rho\sigma\delta}R_{\lambda\rho\sigma\delta})
-4\nabla_{\lambda}\nabla_{\rho}(RR^{\lambda~~\rho}_{(\mu\nu)}),\cr
9)&&:R^{\mu\nu}_{~~\lambda\rho}R^{\lambda\rho}_{~~\sigma\delta}R^{\sigma\delta}_{~~\mu\nu}
\Rightarrow\cr
&&E^{(9)}_{\mu\nu}=-\ft12g_{\mu\nu}R^{\tau\eta}_{~~\lambda\rho}R^{\lambda\rho}_{~~\sigma\delta}R^{\sigma\delta}_{~~\tau\eta}
+\ft32R^{\tau}_{~\mu\lambda\rho}R^{\lambda\rho}_{~~\sigma\delta}R^{\sigma\delta}_{~~\tau\nu}
+\ft32R^{\tau}_{~\nu\lambda\rho}R^{\lambda\rho}_{~~\sigma\delta}R^{\sigma\delta}_{~~\tau\mu}\cr
&&\qquad\quad +\ft32\nabla^{\sigma}\nabla^{\delta}(R_{\sigma\mu}^{~~\lambda\rho}R_{\lambda\rho\delta\nu})
+\ft32\nabla^{\sigma}\nabla^{\delta}(R_{\sigma\nu}^{~~\lambda\rho}R_{\lambda\rho\delta\mu}),\cr
10)&&:R^{\mu~\nu}_{~\lambda~\rho}R^{\lambda~\rho}_{~\sigma~\delta}R^{\sigma~\delta}_{~\mu~\nu}
\Rightarrow\cr
&&E^{(10)}_{\mu\nu}=-\ft12g_{\mu\nu}R^{\tau~\eta}_{~\lambda~\rho}R^{\lambda~\rho}_{~\sigma~\delta}R^{\sigma~\delta}_{~\tau~\eta}
+\ft32R^{\sigma~\rho}_{~\delta~\tau}R^{\delta~\tau}_{~\lambda~\mu}R^{\lambda}_{~\sigma\nu\rho}
+\ft32R^{\sigma~\rho}_{~\delta~\tau}R^{\delta~\tau}_{~\lambda~\nu}R^{\lambda}_{~\sigma\mu\rho}\cr
&&\qquad\quad-\ft32\nabla_{\delta}\nabla_{\sigma}(R_{\mu}^{~\lambda\sigma\rho}R^{\delta}_{~\lambda\nu\rho})
-\ft32\nabla_{\delta}\nabla_{\sigma}(R_{\nu}^{~\lambda\sigma\rho}R^{\delta}_{~\lambda\mu\rho})
+3\nabla_{\delta}\nabla_{\sigma}(R_{(\mu~\nu)}^{~~\lambda~\rho}R^{\delta~\sigma}_{~\lambda~\rho}).
\end{eqnarray}



\begin{thebibliography}{99}


\bibitem{desjactem}
  S.~Deser, R.~Jackiw and S.~Templeton,
{\it Topologically massive gauge theories},
  Annals Phys.\  {\bf 140} (1982) 372.
  [Erratum-ibid.\  {\bf 185}, 406.1988\ APNYA,281,409
(1988\ APNYA,281,409-449.2000)].

\bibitem{lisost}
W.~Li, W.~Song and A.~Strominger,
{\it Chiral gravity in three dimensions},
JHEP {\bf 0804} (2008) 082, arXiv:0801.4566 [hep-th].


\bibitem{berhohtow}
E.~A.~Bergshoeff, O.~Hohm and P.~K.~Townsend,
{\it Massive gravity in three dimensions},
  Phys.\ Rev.\ Lett.\  {\bf 102}, 201301 (2009),
arXiv:0901.1766 [hep-th].

\bibitem{lpcritical}
  H.~L\"u and C.N.~Pope,
{\it Critical gravity in four dimensions,}
  Phys.\ Rev.\ Lett.\  {\bf 106}, 181302 (2011),
  arXiv:1101.1971 [hep-th].

\bibitem{mald}
  J.~Maldacena,
{\it Einstein gravity from conformal gravity,}
  arXiv:1105.5632 [hep-th].


\bibitem{lopapova}
  H.~L\"u, Y.~Pang, C.~N.~Pope and J.~Vazquez-Poritz,
  {\it AdS and Lifshitz black Holes in conformal and Einstein-Weyl gravities},
Phys.\ Rev.\ D {\bf 86}, 044011 (2012),   arXiv:1204.1062 [hep-th].

\bibitem{Riegert:1984zz}
  R.J.~Riegert,
{\it Birkhoff's theorem in conformal gravity,}
Phys.\ Rev.\ Lett.\  {\bf 53}, 315 (1984).

\bibitem{Liu:2012xn}
  H.-S.~Liu and H.~L\"u,
{\it Charged rotating AdS black hole and its thermodynamics in
conformal gravity,}
arXiv:1212.6264 [hep-th].

\bibitem{Mannheim:1990ya}
  P.D.~Mannheim and D.~Kazanas,
{\it Solutions to the Kerr and Kerr-Newman problems in fourth
order conformal Weyl gravity,}
Phys.\ Rev.\ D {\bf 44}, 417 (1991).


\bibitem{mets}
  R.R. Metsaev,
{\it 6d conformal gravity,}
J. Phys. {\bf A44}, 175402 (2011), arXiv:1012.2079 [hep-th].

\bibitem{lupapo}
H.~L\"u, Y.~Pang and C.N.~Pope, {\it Conformal gravity and
extensions of critical gravity,} Phys.\ Rev.\ D {\bf 84}, 064001
(2011), arXiv:1106.4657 [hep-th].

\bibitem{bcn}
F. Bastianelli, G. Cuoghi and L. Nocetti,
{\it Consistency conditions and trace anomalies in six-dimensions,}
Class. Quant. Grav. {\bf 18}, 793 (2001), hep-th/0007222.

\bibitem{Wald:1993nt}
  R.~M.~Wald,
{\it Black hole entropy is the Noether charge},
Phys.\ Rev.\ D {\bf 48}, 3427 (1993), gr-qc/9307038.

\bibitem{Iyer:1994ys}
  V.~Iyer and R.~M.~Wald,
{\it Some properties of Noether charge and a proposal for
dynamical black hole entropy,}
Phys.\ Rev.\ D {\bf 50}, 846 (1994),  gr-qc/9403028.

\bibitem{Ashtekar:1984zz}
  A.~Ashtekar and A.~Magnon,
{\it Asymptotically anti-de Sitter space-times,}
Class.\ Quant.\ Grav.\  {\bf 1}, L39 (1984).

\bibitem{ashdas}
A.~Ashtekar and S.~Das,
{\it Asymptotically Anti-de Sitter space-times:
  Conserved quantities,}  Class.\ Quant.\ Grav.\  {\bf 17}, L17 (2000),
hep-th/9911230.

\bibitem{Bhmrz}
  E.~A.~Bergshoeff, S.~de Haan, W.~Merbis, J.~Rosseel and T.~Zojer,
  {\it On three-dimensional tricritical gravity,}
Phys.\ Rev.\ D {\bf 86}, 064037 (2012),arXiv:1206.3089 [hep-th].

\bibitem{Nut}
T.~Nutma,
  {\it Polycritical gravities,}  Phys.\ Rev.\ D {\bf 85}, 124040 (2012),
arXiv:1203.5338 [hep-th].

\bibitem{knv}
A.~Kleinschmidt, T.~Nutma and A.~Virmani,
{\it On unitary subsectors of polycritical gravities,}
arXiv:1206.7095 [hep-th].

\bibitem{ap}
L.~Apolo and M.~Porrati,
{\it Nonlinear dynamics of parity-even tricritical gravity in three
and four dimensions,} JHEP {\bf 1208}, 051 (2012),
arXiv:1206.5231 [hep-th].

\bibitem{koneto}
C.~N.~Kozameh, E.~T.~Newman and K.~P.~Tod,
{\it Conformal Einstein spaces,} Gen.\ Rel.\ Grav.\  {\bf 17}, 343 (1985).

\bibitem{gonu}
  A.~R.~Gover and P.~Nurowski,
  {\it Obstructions to conformally Einstein metrics in $n$ dimensions,}
math/0405304 [math-dg].

\bibitem{oliray} J.~Oliva and S.~Ray,
{\it Classification of six derivative Lagrangians of gravity and
static spherically symmetric solutions},
Phys.\ Rev.\ D {\bf 82}, 124030 (2010), arXiv:1004.0737 [gr-qc].

\bibitem{Kastor:2009wy}
  D.~Kastor, S.~Ray and J.~Traschen,
  {\it Enthalpy and the Mechanics of AdS Black Holes,}
  Class.\ Quant.\ Grav.\  {\bf 26}, 195011 (2009), arXiv:0904.2765 [hep-th].

\bibitem{Cvetic:2010jb}
  M.~Cvetic, G.~W.~Gibbons, D.~Kubiznak and C.~N.~Pope,
  {\it Black Hole Enthalpy and an Entropy Inequality for the Thermodynamic Volume,}
  Phys.\ Rev.\ D {\bf 84}, 024037 (2011), arXiv:1012.2888 [hep-th].

\bibitem{Dolan:2013ft}
  B.~P.~Dolan, D.~Kastor, D.~Kubiznak, R.~B.~Mann and J.~Traschen,
  {\it Thermodynamic Volumes and Isoperimetric Inequalities for de Sitter Black Holes,}
  arXiv:1301.5926 [hep-th].


\bibitem{Lu:2012ag}
  H.~L\"u and Z.-L.~Wang,
{\it Exact Green's function and Fermi surfaces from conformal gravity,}
  Phys.\ Lett.\ B {\bf 718}, 1536 (2013), arXiv:1210.4560 [hep-th].

\bibitem{Li:2012gh}
  J.~Li, H.~-S.~Liu, H.~L\"u and Z.-L.~Wang,
{\it Fermi surfaces and analytic Green's functions from conformal gravity,}
  arXiv:1210.5000 [hep-th].

\bibitem{lmv}
  H.~Liu, J.~McGreevy and D.~Vegh,
{\it Non-Fermi liquids from holography,}
  Phys.\ Rev.\ D {\bf 83}, 065029 (2011), arXiv:0903.2477 [hep-th].

\bibitem{flmv}
  T.~Faulkner, H.~Liu, J.~McGreevy and D.~Vegh,
{\it Emergent quantum criticality, Fermi surfaces, and AdS$_2$,}
  Phys.\ Rev.\ D {\bf 83}, 125002 (2011)
arXiv:0907.2694 [hep-th].


\bibitem{Duff:1996cf}
  M.~J.~Duff, H.~Lu and C.~N.~Pope,
  {\it Heterotic phase transitions and singularities of the gauge
dyonic string,}  Phys.\ Lett.\ B {\bf 378}, 101 (1996),
hep-th/9603037.

\bibitem{Lu:2012am}
  H.~L\"u and Z.-L.~Wang,
{\it Supersymmetric asymptotic AdS and Lifshitz solutions in
Einstein-Weyl and conformal supergravities,}
  JHEP {\bf 1208}, 012 (2012), arXiv:1205.2092 [hep-th].


\bibitem{Okuyama:2005fg}
  N.~Okuyama and J.~i.~Koga,
  {\it Asymptotically anti de Sitter spacetimes and conserved quantities in
  higher curvature gravitational theories,}
  Phys.\ Rev.\  D {\bf 71}, 084009 (2005)
  arXiv:hep-th/0501044.

\bibitem{Pang:2011cs}
Y.~Pang,
{\it Brief note on AMD conserved quantities in quadratic
curvature theories,}
  Phys.\ Rev.\  D {\bf 83}, 087501 (2011),
  arXiv:1101.4267 [hep-th].

\bibitem{Oliva:2010eb}
  J.~Oliva and S.~Ray,
 {\it A new cubic theory of gravity in five dimensions: Black hole, Birkhoff's theorem and C-function,}
  Class.\ Quant.\ Grav.\  {\bf 27}, 225002 (2010), arXiv:1003.4773 [gr-qc].

\bibitem{Myers:2010ru}
  R.~C.~Myers and B.~Robinson,
{\it Black holes in quasi-topological gravity,}
JHEP {\bf 1008}, 067 (2010), arXiv:1003.5357 [gr-qc].

\bibitem{Chen:2011in}
  Y.X.~Chen, H.~L\"u and K.N.~Shao,
{\it Linearized modes in extended and critical gravities,}
arXiv:1108.5184 [hep-th].






\end{thebibliography}
\end{document}